\documentclass[12pt]{article}           
\long\def\new#1\endnew{{\bf #1}}    
\def\preprint{HUB-EP-00/13\\TUW--00-07}       
\def\finished{February 2000}
\def\archive {hep-th/0002240}           

\def\title{Complete classification of reflexive \\[5mm] 
           polyhedra in four dimensions}

\long\def\abstract{ 
Four dimensional reflexive polyhedra encode the data for smooth Calabi--Yau
threefolds that are hypersurfaces in toric varieties, and have important 
applications both in perturbative and in non-perturbative string theory.
We describe how we obtained all 473,800,776 reflexive polyhedra
that exist in four dimensions and the 30,108 distinct pairs of Hodge numbers 
of the resulting Calabi--Yau manifolds. As a by-product we show that all these
spaces (and hence the corresponding string vacua)
are connected via a chain of singular transitions. 
\del
In addition to the 308
maximal reflexive polyhedra that contain all others as subpolyhedra on 
sublattices, there are 32 reflexive polyhedra that are maximal on a fixed
lattice.
\enddel
}

\def\CY{Calabi--Yau}

\def\ipo{\hbox{\bf 0}} 
\def\rp{reflexive polyhedron}
\def\coa{_{\rm coarsest}}
\def\fin{_{\rm finest}}
\def\q{{\mathbf q}}
\def\Dq{\D(\q)}
\def\cont#1{\mathop{\vtop{\ialign{##\crcr $\hfil\displaystyle
                     {#1}\hfil$\crcr\noalign{\kern3 pt \nointerlineskip}
                     $\bracelu\leaders\vrule\hfill\leaders\vrule\hfill
                     \braceru$\crcr\noalign{\kern3 pt }}}}\limits}

\def\bpic{\begin{picture}} \def\epic{\end{picture}} \thicklines
\def\lab#1)#2#3{\put#1){\makebox(0,0)[#2]{\small #3}}}
\def\putlin#1,#2,#3,#4,#5){\put#1,#2){\line(#3,#4){#5}}} 
\def\putvec#1,#2,#3,#4,#5){\put#1,#2){\vector(#3,#4){#5}}}

\newcount{\vmul} \newcount{\hdiv} 
\newcount{\wi}   \newcount{\he}   
\newcount{\hq}   \newcount{\vq}   
\newcount{\hoff} \newcount{\auxc} \newcounter{figco}   \def\npt{\circle*{2}}

\def\vlline{\put(-3,0){\line(1,0)6}}      

\def\putvm#1{\mbox{\bpic(0,0)\funit=1pt\vlline\epic}}   

\def\Vpt#1,#2){
\hq=#1\multiply\hq by 2   
\vq=#2\multiply\vq by 2 \advance\vq by #1
\divide\hq by\hdiv     \multiply\vq by\vmul  
\put(-\hq,\vq){\npt}}
\def\Vplo#1{\vbox{\hdiv=1\vmul=1 \figsca \auxc=\he \multiply\auxc by\vmul
    \hoff=\wi
\stepcounter{figco}\message{[Fig. \arabic{figco}}
    \begin{center}\let\.=\Vpt \bpic(\wi,\auxc)(-\hoff,0) \figlab #1 
      \put(-\hoff,0){\vector(1,0){\auxc}}
      \put(-\hoff,0){\vector(0,0){\auxc}}
      \epic \\[5mm]
    Fig. \arabic{figco}: \figcap \end{center}} \vfil \message{]}}
\def\figsca{\unitlength=1.1pt \wi=500 \he=400} \let\funit=\unitlength

\input epsf

\usepackage{amssymb,latexsym}                   \catcode`\"=\active \let"=\"

\def\ifundefined#1{\expandafter\ifx\csname#1\endcsname\relax}
\def\bye{\end{document}}   

\def\HS#1 {\hspace*{#1pt}} \def\VS#1 {\vspace*{#1pt}} \long\def\del#1\enddel{} 
\def\BC{\begin{center}}    
\def\EC{\end{center}}

\let\0=\over      \let\1=\vec      \def\2{{1\over2}}    \let\3=\ss
\let\4=\underline \let\5=\overline \let\6=\partial      \def\7#1{{#1}\llap{/}}
\def\8#1{{\textstyle{#1}}}         \def\9#1{{\ifmmode{\pmb{#1}}\else\bf#1\fi}}

          \def\({\left(}       \def\){\right)}

\def\eeql#1 {\label{#1}\eeq}        
\def\beq{\begin{equation}}      \def\eeq{\end{equation}}        
\def\bea{\begin{eqnarray}}      \def\eea{\end{eqnarray}}

\let\and=\wedge

\let\bra=\langle        \let\ket=\rangle        \def\<#1\>{\bra #1 \ket}
\let\ni=\noindent

\def\rel#1 #2{\buildrel #1 \over {#2}}  
\def\fnote#1#2{\begingroup\def\thefootnote{#1}\footnote{#2}
                \addtocounter{footnote}{-1}\endgroup}   

\def\subdef#1{\gdef\globalColor##1{##1}}      
      \subdef{Black}

         \let\th=\theta  
   \let\l=\lambda

             \let\D=\Delta

\def\IR{{\mathbb R}}  \def\IP{{\mathbb P}} 
   
\def\IZ{{\mathbb Z}}

\def\plb#1 #2 {Phys. Lett. {\bf B#1} #2 }
\def\phr#1 #2 {Phys. Rep. {\bf  #1} #2 }        
\def\npb#1 #2 {Nucl. Phys. {\bf B#1} #2 }
\def\aph#1 #2 {Ann. Phys. {\bf #1} #2 }         
\def\jmp#1 #2 {J. Math. Phys. {\bf #1} #2 }
\def\jgp#1 #2 {J. Geom. Phys. {\bf #1} #2 }
\def\prd#1 #2 {Phys. Rev. {\bf D#1} #2 }
\def\prl#1 #2 {Phys. Rev. Lett. {\bf #1} #2 }
\def\rmp#1 #2 {Rev. Mod. Phys.  {\bf #1} #2 }
\def\zpc#1 {Z. Phys. {\bf #1C} }
\def\cmp#1 #2 {Commun. Math. Phys. {\bf #1} #2 }
\def\cqg#1 #2 {Class.Quant.Grav. {\bf #1} #2 }
\def\mpl#1 {Mod. Phys. Lett. {\bf A#1} }
\def\cpc#1 {Computer Phys. Commun. {\bf #1} }   
\def\ijmp#1 {Int. J. Mod. Phys. {\bf A#1} }
\def\ijmpC#1 {Int. J. Mod. Phys. {\bf C#1} }

\def\BP{\begin{picture}} \def\EP{\end{picture}}         
\newcounter{TRefNX} \let\OLDcite=\cite  \makeatletter
\def\makeTRefs#1{\@for  \NewTRef:=#1\do{\global\makeTRef{\NewTRef}}}
\def\makeTRef#1{\ifundefined{TRef#1}\stepcounter{TRefNX}%
\expandafter\xdef\csname TRef#1\endcsname{\theTRefNX}\fi}\makeatother
\def\NEWcite#1{\makeTRefs{#1}\OLDcite{#1}}  

\ifundefined{draftmode} {} \else        \let\cite=\NEWcite
\newcount\HOUR  \HOUR=\the\time \divide\HOUR by 60      \multiply \HOUR by 60 
\newcount\MIN   \MIN=\the\time  \advance\MIN by -\HOUR  \divide\HOUR by 60
\ifnum  \MIN>9  \def\printTIME{{\it\the\HOUR\,:\,\the\MIN}}
        \else   \def\printTIME{{\it\the\HOUR\,:\,0\the\MIN}} \fi 

\def\LLab#1{\BP(0,0)\unitlength=1mm\put(-12,.5){\makebox(0,0)[cr]{\small #1
        \rlap{$_{_{\makeatletter\csname TRef#1\endcsname\makeatother}}$}}}\EP}
\fi

\begin{document}

\vspace*{-20mm}\begin{flushright}       \archive\\\preprint  \end{flushright}
\vspace*{-5mm}
\begin{center}  {\huge\bf       \title  }\end{center}
\begin{center} \vskip 3mm
        Maximilian KREUZER\fnote{*}{e-mail: maximilian.kreuzer@tuwien.ac.at}
\\[3mm]
        Institut f"ur Theoretische Physik, Technische Universit"at Wien\\
        Wiedner Hauptstra\3e 8--10, A-1040 Wien, Austria
\\[4mm]                       and
\\[4mm] Harald SKARKE\fnote{\#}{e-mail: skarke@physik.hu-berlin.de}
\\[3mm] Humboldt-Universit\"at, Institut f\"ur Physik, QFT, \\
        Invalidenstra\ss e 110, D-10115 Berlin, Germany
        
\vfill\vfill    {\bf ABSTRACT } \end{center}\vspace{-2mm}    \abstract

\vfill\vfill \noindent \preprint\\[1pt] \finished \vfill
\thispagestyle{empty} \newpage
\pagestyle{plain} 

\newpage
\setcounter{page}{1}

\section{Introduction and basic definitions}

Calabi--Yau manifolds provide some of the most beautiful examples of the
interplay between theoretical physics and what was previously considered to be
pure mathematics.
With the discovery of the relevance of Calabi--Yau manifolds to string
compactifications \cite{CHSW} it became necessary for physicists to get
acquainted with techniques of algebraic geometry, but soon physicists were
able to formulate conjectures that became the focus of mathematicians'
interest, such as mirror symmetry \cite{dix,lvw} and its application to the
problem of counting rational curves on Calabi--Yau manifolds 
\cite{exact1,exact2}.
Conversely, mathematical developments such as the introduction of homogeneous
coordinates for toric varieties \cite{Cox} were almost immediately exploited
for applications in string theory.
While physicists found that the space of Calabi--Yau hypersurfaces in weighted
projective spaces is almost perfectly symmetric under mirror symmetry 
\cite{CLS,KlS,nms}, and also provided the first explicit
constructions of mirror manifolds \cite{gp,BH}, it was the mathematician 
Batyrev \cite{bat} who provided the framework necessary for making this 
symmetry complete and explicit, namely the description of Calabi--Yau
manifolds that are hypersurfaces in toric varieties in terms of reflexive
polyhedra.

Mirror symmetry is not the only reason why hypersurfaces and, more generally, 
complete intersections in toric varieties play a particularly prominent role
among all Calabi--Yau manifolds.
They constitute nearly all of the examples used in the physics
literature, and they are also precisely the spaces occurring in the
geometrical phase of Witten's gauged linear sigma model \cite{Wph}, which
not only explains the connection between Landau--Ginzburg models and
Calabi--Yau compactifications, but is also relevant to the discussion of
topology changing physical processes \cite{AGM1,AGM2}.
In the context of string dualities and non-perturbative string physics,
reflexive polyhedra can be used, for example, for directly finding enhanced
gauge groups \cite{CF}.

Another point where physics and mathematics interact very strongly is the
question of connectedness of the moduli space of Calabi--Yau manifolds.
While this question was originally asked by a mathematician \cite{Reid2},
it was later found that physical processes \cite{Str,gms} may interpolate
between the corresponding different branches of the moduli space.
Here, again, an important role is played by reflexive polyhedra: 
They were used to demonstrate the connectedness of the space of Calabi--Yau
hypersurfaces in weighted projective spaces \cite{CGGK,ACJM}.
In terms of polyhedra, two Calabi--Yau manifolds are connected if one of the
corresponding polytopes is a subpolyhedron of the other or, more generally, if
between the two polyhedra there is a chain of polyhedra that are mutually
connected in this way.

For all of these reasons, a complete classification of reflexive polyhedra and
thus of smooth Calabi--Yau manifolds that are hypersurfaces in toric
varieties, is an important task.
Until recently, the complete set of reflexive polyhedra was known only for the
case of two dimensions, corresponding to one dimensional Calabi--Yau
manifolds, i.e. elliptic curves.
To change this situation, we developed an algorithm for the classification in
arbitrary dimensions \cite{crp,wtc} and, as a first step, used it to find all
4319 reflexive polyhedra in three dimensions \cite{c3d}, which give rise to K3
manifolds.
In the present work we present the results of a complete classification in
four dimensions, and as a side result we can state that the corresponding
moduli space of Calabi--Yau threefolds is connected (this was almost, but not
quite completely shown in \cite{web}).
We plan to make our complete results accessible at our web site \cite{KScy}.

In the remainder of this introductory section we give some of the necessary
defintitions. 
In section two we briefly outline our classification algorithm and in section
three we discuss the modifications and tricks that were necessary to make this
algorithm work in four dimensions.
Finally we present our results, both at the level of polyhedra and in terms of
the geometry of Calabi--Yau threefolds.

Reflexive polyhedra are defined in relation to some lattice 
$M\simeq \IZ^n$ or its dual lattice $N$, and the underlying real vector spaces
$M_\IR$ and $N_\IR$.
They always have the origin $\ipo$ in their interiors;
we will denote this property as the `interior point property' or `IP property'.
Given such an `IP polytope', the
dual (or polar) polytope $\D^*\subset N_\IR$ of $\D\subset M_\IR$ is defined
as
\beq 
\D^*=\{y\in N_{\IR}: ~~~\<y,x\>\ge -1 ~~~\forall x\in \D\},  
\eeql{dual}
where $\<y,x\>$ is the duality pairing between $y\in N_{\IR}$ and $x\in M_\IR$.
Because of the convexity of $\D$, $(\D^*)^*=\D$.

Given the lattice $M$, a lattice (or integer) polyhedron is a polyhedron on 
$M_{\IR}$ whose vertices lie in $M$.
A lattice polyhedron $\D\subset M_{\IR}$ is called reflexive if its dual 
$\D^*\subset N_{\IR}$ is a lattice polyhedron w.r.t. $N$.
Of course the same definitions are valid upon interchanging $M$ with $N$ and
$M_{\IR}$ with $N_{\IR}$.

\section{Outline of the algorithm}

The basic idea of our algorithm is to construct a finite set $S$ of ``maximal''
polyhedra such that every reflexive polyhedron is a subpolyhedron of some 
polyhedron in $S$. 
Given $S$, all that is left to do is to construct all subpolyhedra of all
polyhedra in $S$ and check for reflexivity.
By duality, the set $S^*$ of polyhedra dual to the ones in $S$ has the
property that any reflexive polyhedron contains at least one element of $S^*$
as a subpolyhedron, so these objects are ``minimal'' in a certain sense.
More precisely, we can give the following definition of minimality:
A minimal polyhedron $\nabla\subset N_{\IR}$ is 
defined by the following properties:\\[3mm]
1. $\nabla$ has the IP property.\\
2. If we remove one of the vertices of $\nabla$, the convex hull of the
  remaining vertices of $\nabla$ does not have the IP property.\\[3mm]
It is easily checked that any \rp\ $\D^*$ contains at least one minimal 
polyhedron. 
As we showed in \cite{crp}, $\nabla$ must either be a simplex or contain
lower dimensional simplices with the origin in their respective interiors such
that any vertex of $\nabla$ belongs to at least one of these simplices.
This leads to a coarse classification with respect to the number of vertices
and the simplices they belong to: 
In two dimensions, the only possibilities are the triangle $V_1V_2V_3$
and the parallelogram $V_1V_2V_1'V_2'$, where  $V_1V_2$ and $V_1'V_2'$ are
one dimensional simplices (line segments) with the origin in their respective 
interiors.
In a shorthand notation, 
where a $p$-simplex is denoted by the number $p+1$ of its vertices,
this result may be summarized as $\{$3; 2+2$\}$.
In the same notation the result for three dimensions can be summarized
as $\{$4; 3+2, $\cont{\hbox{3+3}}$; 2+2+2$\}$ where the underlining symbol 
indicates a vertex occurring in both triangles 
(an Egyptian pyramid is an example of such a configuration).
In the four dimensional case of our present study, the possibilities can be
summarized as 
\beq \hbox{
$\{$5; 4+2, 3+3, $\cont{\hbox{4+3}}$, $\cont{\cont{\hbox{4+4}}}$;
 3+2+2, $\cont{\hbox{3+3}}$+2, 
 $\cont{\hbox{3+3}\hskip -3pt}\hskip -2pt\cont{\hbox{~+3}}$; 2+2+2+2$\}$,}
\eeq
where the three triangles in 
$\cont{\hbox{3+3}\hskip -3pt}\hskip -2pt\cont{\hbox{~+3}}$
have one common vertex and configurations with the same number of vertices 
are separated by commas.

The fact that a simplex spanned by vertices $V_i$ contains the origin in its
interior is equivalent to the condition that there exist positive real numbers
(weights) $q_i$ such that $\sum q_i V_i=0$. 
We call a set of weights for a simplex a weight system and the collection of
weight systems for a minimal polyhedron (with $q_i^{(j)}=0$ if $V_i$ does
not belong to the j'th simplex) a combined weight system or CWS.
Perhaps the most important feature of these combined weight systems is the
fact that they allow us to reconstruct $\nabla^*$ and hence $\nabla$ in a
simple way: 
If $\nabla$ has the vertices $V_1,\ldots,V_k$, consider the map 
\beq M_\IR\to \IR^k, \quad X\to{\mathbf x}=(x_1,\ldots,x_k) \hbox{ with }
     x_i=\<V_i,X\>.   \eeq
It is easily checked that this map defines an embedding such that the image 
of $M_\IR$ is the subspace defined by $\sum_i q_i^{(j)} x_i=0$ $\forall j$.
Moreover, $\nabla^*$ is isomorphic to the polyhedron defined in this subspace 
by $x_i\ge -1$ for $i=1,\ldots,k$.

Looking for reflexive polytopes, we obviously require $\nabla$ to have the
following two additional properties:\\[3mm]
3. $\nabla$ is a lattice polyhedron.\\
4. The convex hull of $\nabla^*\cap M$ has the IP property.\\[3mm]
3. implies that the $q_i$ can be chosen as $n_i/d$, where the
$n_i$ are integers without a common divisor and $d=\sum_i n_i$.
In this case there is a unique lattice $N\coa$ that is generated by the
vertices of $\nabla$ such that any lattice $N$ on wich $\nabla$ is integer is
a refinement of $N\coa$.
If we denote the lattice dual to $N\coa$ as $M\fin$, then it is easily checked
that the image of $M\fin$ under our embedding map is nothing but the set of
integer $(x_1,\ldots,x_k)$ fulfilling $\sum_i n_i^{(j)} x_i=0$ $\forall j$.
If 4. is fulfilled by any lattice $M$, it must also be fulfilled by $M\fin$,
so for analysing which weight systems play a role it is sufficient to restrict
our attention to $M\fin$.
Among the consequences of the embedding map is the fact that a CWS $\q$
uniquely determines $\nabla$ and $M\fin$.
Given $\q$, we define $\Dq$ as the convex hull of $\nabla^*\cap M\fin$ and 
say that $\q$ has the IP property if $\Dq$ has the IP property.
It is not hard to check that a CWS can have the IP property
only if each of the single weight systems occurring in it has the IP property.
As shown in \cite{wtc}, there is only a finite number of weight systems $\q$ 
with a given number of weights such that $\Dq$ has the IP property.
For 2 weights, there is just the system $(n_1,n_2)=(1,1)$, with 3 weights
there are the systems $(1,1,1)$, $(1,1,2)$ and $(1,2,3)$ and with 4 and 5
weights there are 95 and 184026 weight systems, respectively.
Plugging them into the simplex structures and checking for 4., one gets
a total of 201346 CWS relevant to four dimensional
reflexive polytopes.

One way of finding all reflexive polyhedra in four dimensions would be to
find all IP subpolyhedra $\D$ of $\Dq$ for the above 201346
CWS and check whether there is a sublattice of $M\fin$
on which $\D$ is reflexive.
The question of whether a given polytope $\D\subset M_\IR$ can be reflexive
on any lattice (and if so, on which lattices) may be approached in the
following way:
If $\D$ has $n_V$ vertices and $n_F$ facets (a facet being a codimension 1 
face), the dual polytope $\D^*$ has $n_V$ facets and $n_F$ vertices.
In this situation there is not only a coarsest lattice $N\coa$ on which
$\D^*$ can be a lattice polytope (generated by the vertices of $\D^*$), but 
also a
coarsest lattice $M\coa$ generated by the vertices of $\D$, as well as
the dual lattices $N\fin$ and $M\fin$, respectively.
We define the vertex pairing matrix $X$ as the $n_F\times n_V$
matrix whose entries are $X_{ij}=\<\bar V_i,V_j\>$, where $\bar V_i$ and
$V_j$ are the vertices of $\D^*$ and $\D$, respectively.
$X_{ij}$ will be $-1$ whenever $V_j$ lies on the i'th facet, and the rank of
$X$ is just the dimension $n$ of $M_\IR$.
If there is any lattice such that $\D$ is reflexive on it, then all entries of
$X$ must be integer.
Conversely, if $X$ is integer, then $M\coa$ is a sublattice of $M\fin$ and 
all lattices on which $\D$ is reflexive can be found as follows:
By recombining the lines and columns of $X$ in the style of Gauss' algorithm
and keeping track of the $GL(\IZ)$ transformation matrices used in the
process, we may decompose $X$ as $X=W\cdot D\cdot U$,
where $W$ is a $GL(n_F,\IZ)$ matrix, $U$
is a $GL(n_V,\IZ)$ matrix and $D$ is an $n_F\times n_V$
matrix such that the first $n$ diagonal elements are positive integers whereas
all other elements are zero.
In the same way as we defined an embedding of $M_\IR$ in $\IR^k$,
we now define an embedding in $\IR^{n_F}$ such that
$M_{\rm finest}$ is isomorphic to the sublattice of $\IZ^{n_F}$ 
determined by the linear relations among the $\bar V_i$.
In this context the $X_{ij}$ are just the embedding coordinates of the $V_j$.
The $GL(n_F,\IZ)$ matrix $W$ effects a change of coordinates in 
$\IZ^{n_F}$ so that $\D$ now lies in the lattice spanned by the first $d$ 
coordinates, i.e. we can interpret the first $n$ entries of the j'th column 
of $D\cdot U$ as coordinates of $V_j$ on $M_{\rm finest}$.
Similarly, the lines of $W\cdot D$ provide coordinates for the vertices of 
$\D^*$ on $N_{\rm finest}$, whereas $U$ and $W$ give the corresponding 
coordinates on the coarsest possible lattices.
The intermediate lattices are in one-to-one correspondence to subgroups
of the finite lattice quotient $M\fin/M\coa$ and can be found by decomposing 
$D$ into a product of triangular matrices as described in \cite{c3d,ams}.

In principle we now have a complete algorithm. 
It turns out, however, that we need far less than the above mentioned 201346 
CWS for our classification scheme if we choose suitable refinements of our
original definition of minimality.
While there are several possibilities of doing this (see \cite{crp,c3d,ams}),
we will mention only the one that is most powerful for the case of four
dimensions considered here.
We call a polytope $\D\subset M_\IR$ r-maximal if $\D$ is reflexive
w.r.t. the lattice $M$ but is not contained in any other polytope that is
reflexive w.r.t. $M$.
The duals $\D^*$ of these objects are called r-minimal; these are the
reflexive polytopes that have no reflexive subpolytopes.
As shown in \cite{wtc}, for $n\le 4$ $\Dq$ is reflexive whenever it has the 
IP property.
In this case we define a CWS to be r-minimal if $\Dq$ is r-maximal.
It is not hard to see that every reflexive polytope must be a subpolytope 
of $\Dq$ on some lattice $M\subseteq M\fin$, where $\q$ is an r-minimal CWS.
With our algorithms it is easy to check for r-minimality by checking whether
$(\Dq)^*$ contains a reflexive proper subpolyhedron.
In this way we found that only 308 of the 201346 CWS are r-minimal
(these 308 CWS are listed in appendix A).
Nevertheless, as we will see, there are further 
 polytopes that are r-maximal with respect to 
proper sublattices of $M\fin$.

Thus we can find all reflexive polytopes in four dimensions by finding all
IP subpolytopes $\D$ of the 308 $\Dq$ with r-minimal $\q$ and applying our 
method for identifying all lattices on which $\D$ is reflexive.

\section{Improvements of our basic algorithm}

For the classification of reflexive polyhedra in three dimensions \cite{c3d}
we used, more or less, an implementation of the algorithm outlined in the
previous section already. 
Our programs, however, would not have worked for the four dimensional case
because everything that can go wrong in large computer calculations would have
gone wrong: Within a rather short time we would have run out of RAM and disk
space, would have encountered numerical overflows, and even without these
problems we would not have been able to obtain our results within a reasonable
time.
To understand where these problems came from and how we eventually managed to
solve them, it is necessary to go into some technical details of our approach.

One of the main parts of our package is a set of routines for analysing a
polytope $\D$ given in terms of the lattice points whose convex hull it
represents with respect to the following questions:\\
What are the vertices of $\D$?\\
What are the equations for the bounding hyperplanes of $\D$?\\
Is $\D$ IP?\\
Is $\D$ reflexive?\\
The basic form of our algorithm is as follows:
Fist one looks among the given points for a set of points that span a simplex
$S=\D_0$ of the full dimension.
By imposing extremality conditions on the coordinates of these points, one can
choose these points in such a way that they are vertices $V_1,\ldots,V_{n+1}$
of $\D$.
Then one checks for every bounding hyperplane $H$ of $\D_0$ if there are
still points of $\D$ outside $H$.
If this is the case, one picks one of these
points, again in such a way that it is a vertex ($V_{n+2}$) of $\D$.
There are new bounding hyperplanes connecting $V_{n+2}$ with vertices of $H$ 
(which is subsequently omitted from the list of bounding hyperplanes).
Proceeding in this way one obtains successively larger polytopes $\D_i$
with vertices $V_1,\ldots,V_{n+1+i}$ whose bounding hyperplanes are known.
Finally, if for some $k$ no point of $\D$ lies outside of $\D_k$, we know that 
$\D=\D_k$ and we have found all vertices and defining equations of $\D$.

In this form, there are the following problems: 
For every new hyperplane, we have to calculate an equation from $n$ points. 
The corresponding linear algebra is time consuming and prone to overflows.
Besides, for four dimensional polytopes, connecting every new vertex with 
every subset of $3$ points of an old hyperplane $H$ with $k$ vertices 
introduces a combinatorial factor of $k\choose 3$
(this was no problem in the three dimensional case with $k\choose 2$ and
usually moderate $k$).
Therefore it is desirable to connect the new vertex only with all codimension
2 faces of $H$. 
At first sight this seems to require even more linear algebra, but this can be
evaded with the following trick:
To every hyperplane $H$ we assign a bit pattern with the $i$'th bit equal to 1
if $V_i$ lies on $H$ and equal to 0 otherwise.
Intersecting hyperplanes then corresponds to bitwise `logical and'
operations, and this type of operation is provided by the operator \& in the C
programming language. 
Among such bit patterns (`unsigned integers' in C) there is a natural partial
ordering defined by the bitwise ordering $0<1$.
In order to find the facets of $H$, one can proceed in the following way:
Among all the bit patterns coming from `bitwise intersections' of $H$ and all
other hyperplanes, the facets of $H$ correspond to the maximal objects with 
respect to the partial ordering.
Thus it is sufficient to connect the new vertex only with the vertices encoded
in the corresponding unsigned integers.

Even the remaining linear algebra can be avoided in the following way:
Every codimension 2 face is the intersection of two facets given by equations 
\beq E^{(i)}(x)= \sum_{j=1}^n a_j^{(i)}x_j - c^{(i)}=0,\qquad i=1,2. \eeq
Any new hyperplane passing through the codimension 2 face must be described by
an equation of the form
\beq E(x)=\l_1E^{(1)}(x)+\l_2E^{(2)}(x)=0, \eeq
so the equation for the new facet passing through the vertex $V$ is just
given (up to scaling) by $\l_1=E^{(2)}(V)$, $\l_2=-E^{(1)}(V)$.
In this way we have replaced the solution of a system of $n$ linear equations
by the evaluation of two expressions.

A further source for saving calculation time comes from the fact that in our
classification scheme the polytopes are generated by starting with a small
number of objects and obtaining all others by dropping vertices from the sets
of lattice points of polyhedra that have already been analysed.
If we obtain a polytope $\tilde \D$ from a polytope $\D$ by dropping the
vertex $V$, then we can make use of the following facts:
All vertices of $\D$ except for $V$ are vertices of 
$\tilde \D$, and all facets of $\D$ not containing $V$ 
are again facets of $\tilde \D$.
Thus we already have the data for $\tilde \D$, except for a `hole' that is
bounded by the codimension 2 faces that are intersections between facets of
$\D$ containing $V$ and other facets.
Then one can first `close the hole', i.e. find the data for the polytope $\D'$
which is the convex hull of the vertices of $\D$ except $V$ and then proceed
to find the remaining vertices and facets of $\tilde \D$ as in our basic 
algorithm.
`Closing the hole' can be done in the following way: 
First one connects the codimension 2 faces bounding the hole with the other 
vertices along the hole in the way described above.
Then one checks whether the resulting collection of facets of $\D'$ is already
complete.
This can be achieved with an algorithm that relies again on intersections of
bit patterns and that returns the vertices belonging to the missing facets.
The missing facets can be found by connecting the corresponding vertices.

Even for the smallest of our r-maximal polytopes (the one with 47 lattice 
points), each of the above tricks (using incidences to find codimension 2 
faces, getting new equations from old ones and using the data of previously 
analysed polytopes) improves calculation time by a factor roughly of the order
of two.
Together they yield a factor of eight, and for larger polytopes the 
improvement is probably even much better.
This takes care of the problems with calculation time and some of the
overflows.
Overflows occurring at other points in our programs could be
circumvented by improving basic routines of linear algebra, for instance by 
rearranging lines and columns before and during the application of Gauss-type 
algorithms, or simply by using  double precision integers (64 bit) for 
certain routines.

This still leaves us with RAM and disk space problems. 
Clearly the first thing to do is to compress our data as far as possible.
A factor of more than two can be gained immediately by making use of mirror
symmetry: For every mirror pair, we kept only the smaller of the two
polytopes, together with information on which of them had already been found.
In order to be able to check efficiently if a polyhedron was new it was 
essential to define a normal form 
based on intrisic properties of the polytope \cite{c3d} and an ordering 
which allowed 
us to store the polyhedra in a sorted list that could be searched by 
bisection. 
The coordinates in the normal form
were then shifted to non-negative values and interpreted as
digits of a number in a sufficiently large basis. 
This number was stored in binary format, 
so that we could put an average of 3 coordinates on one byte. 
With these methods we needed about 20 bytes to encode  
a mirror pair of reflexive polyhedra, reducing the size of our
data to approximately 4.5 GB which fit easily 
on the harddisks we were using but not into the RAM of our computers.
This meant that whenever our programs found a reflexive polytope, it had (at
least in principle) to read frome the hard disk to see whether this polytope
had been found before.
To avoid the enormous time consumption involved in reading large portions from
the disk, we had to structure our data in the form of a database in such a way
that disk operations were minimized.
We achieved this by keeping approximately every 64'th polytope in the RAM, so
that after a bisecting routine using only data in the RAM the program had to
read only a single block of 64 or less polytopes from the disk.
As a side effect we could also make effective use of the disk cache in this
way, i.e. in most cases even the reading of this block was substituted by a
(much faster) reading from the disk cache.

Another point that required care, in particular in view of the question of
connectedness, was the distinction between polyhedra that are reflexive on the
original lattice $M_{\rm finest}$ of one of the 308 polytopes coming from 
r-minimal CWS (we will call them `OL--polyhedra'), and reflexive polyhedra 
on sublattices.
We chose the following strategy: 
We kept a large list of OL--polyhedra and in addition a small list of 
polyhedra that were not reflexive on the original lattice but on some
sublattice and were not (yet) contained in the list of OL--polyhedra. 
The complete list can then be generated by looking for all possible lattices
for all polytopes of either list.

The two most time consuming tasks in our program are the analysis of a
polyhedron, which can be performed about 2100 
times per second on a PentiumIII/600 and 1400 times on 
an Origin 2000 processor, respectively, and the evaluation
of the normal form, for which the corresponding numbers are 1800 and 900.
The PC is therefore faster by a factor of 3:2 for the first task and by a 
factor of 2:1 for the evaluation of the normal form although everything is done
with pure integer arithmetics. 
This apparent discrepancy is probably explained by
the 64-bit architecture of the Origin 2000: Whereas the time-critical part
of the evaluation of the normal form only uses 32-bit integers, the analysis
of the incidence structures of the polyhedra required 64 bit, since the
vertex number often exceeded 32 (but never 64).
As most of the polyhedra that occurred in our computations were 
not reflexive (in which case no normal form was calculated), the main portion 
of the total CPU time was consumed by polyhedron analysis.

\section{Results}

After more than half a year, using two dual Pentium III/600MHz PCs and 
between 10 and 30 processors on a 64 processor SGI Origin 2000, 
our programs produced the following results:
A list of 473,800,652 reflexive polytopes that are subpolyhedra of the 308 
polytopes we started with on the original lattice (i.e., OL--subpolyhedra), 
and 98 additional polytopes that are subpolytopes of the 308 on a sublattice
but not reflexive on the original lattice.\footnote{
More precisely, whenever we found a polytope only on a sublattice in 
the `big run', we didn't proceed to look for its subpolyhedra. 
In this way we got 70 such polytopes, and in a second run found that they had 
28 subpolytopes that had not yet been found in either list.
}
Among the 473,800,652 OL--polyhedra there are only 124 polytopes whose duals 
are not in the same list.
Looking for different lattices for the OL--polyhedra yields 95 polytopes not
yet in that list, and looking for all possible lattices for the 
other 98 polyhedra leads to 102 polytopes not among the 473,800,652.
The union of the respective sets of 95 and 102 polyhedra consists
precisely of the 124 polytopes that are required to make the large list
mirror symmetric, leading to a total of 473,800,776 reflexive polyhedra.
More precisely, there are 236,879,533 dual pairs and 41,710 self-dual
polyhedra.

At this point it is appropriate to discuss the reliability of our results.
Given the numbers of polyhedra that we found, the complexity of our
programs, and the calculation time of several processor years, one may 
certainly wonder whether our results should be trusted.
The reason why we are sure that they are correct is the following:
While our whole construction is manifestly asymmetric with respect to duality,
starting with `large' polyhedra and getting all others as subpolytopes, the 
result is perfectly symmetric.
Assuming that by some unknown error we had missed precisely 1 polytope, the 
probability that it were self mirror (and thus not seen to be lacking) would 
be around $10^{-4}$; if we had missed 2 polytopes, there would have been a
chance of roughly $10^{-8}$ of both being self mirror and of approximately 
$2\times 10^{-9}$ that these two polytopes constituted a mirror pair.
Clearly one would expect most sources for errors in our results to produce
larger numbers of missing polytopes, making the perfect symmetry of our
results even more unlikely.%
\footnote{Actually, this is what really happened: In the first run we ended up
        with 476 missing mirror polyhedra. It turned out that all of them
        could only come from 3 different r-maximal objects. Moreover, all of
        the 3 respective files were created in the last days of November 99
        on the Origin 2000. It seems that some processor(s) on that machine
        had a problem at that time.
        Repeating the calculation 
        for these 3 files produced exactly all missing mirrors and the desired
        symmetric result.
}
We therefore conclude that the only reasonable explanation for the symmetry 
of our results is their correctness.

As we mentioned, every reflexive polytope is a subpolytope of one of 308
objects either on the same lattice or on some sublattice.
Asking ourselves about a complete list of r-maximal objects (such that every
reflexive polytope is a subpolytope of one of them on the same lattice), it is
clear that we need additional polyhedra because of the 124 polytopes that are
not found on the original lattices.
This was to be expected, since even in three dimensions there is a polytope
that is not of this type \cite{c3d}.
As a matter of fact, even before we finished the classification we could find 
25 additional r-maximal polytopes by considering the original 308 objects on 
different lattices.
It turned out, however, that even taking into account the subpolytopes of
these 25 polyhedra was not sufficient to complete the full list.
Given the fact that all except 124 reflexive polytopes are OL--subpolytopes 
of the 308, it is clear that the search for additonal r-maximal objects can be
restricted to these 124 polytopes.
The result is that these 124 polytopes contain 32 r-maximal polytopes, i.e. in
addition to the $308+25$ polyhedra mentioned above there are 7 further 
r-maximal polytopes, yielding a total of $308+25+7=340$ r-maximal polyhedra in
four dimensions.
Information on these objects is given in appendix A. 
Tables 1--7 contain the 308 CWS that define r-maximal polytopes of the type
$\D(\q)$ as well as information on additional r-maximal polytopes
that we obtained by considering $\D(\q)$ on sublattices of $M\fin$.
It turns out that the corresponding 25 polytopes all occur for `Fermat
type' CWS where $\nabla^*=\D(\q)$.
The 7 last polytopes (more precisely, their r-minimal duals) which are of 
neither of the aforementioned types are listed in table 8
in the normal form coordinates produced by our programs.
Despite their different structure, each of these r-minimal polytopes 
contains a single minimal subpolytope, which is not reflexive and
turns out to be a simplex in each of the 7 cases.
Thus even these polytopes are connected with weight systems in a unique way.
For example, the first polytope in table 8 contains a simplex with the same
weights as the reflexive simplex corresponding to the quintic hypersurface, 
but on a lattice on which it is not reflexive. 
Addition of the vertex $v_1$ turns this simplex into an r-minimal polytope.

It is now possible to state the connectedness of all 
four dimensional polyhedra in the sense described in the introduction:
Connectedness of all polyhedra is now equivalent to connectedness of the 340
r-maximal polytopes.
This could be shown as a by-product of our classification scheme: 
During the determination of the subpolytopes of the 308, we checked that each
r-maximal polytope contained at least one OL-subpolytope that is also an
OL-subpolytope of some other of the 308 objects.
This re-established the connectedness of the 308, which had been shown in
\cite{web} already.
Finally, by considering the OL-subpolytopes of the remaining 32 r-maximal
polytopes, connectedness of all reflexive polyhedra was established.

Before proceeding to an interpretation of our results in terms of the 
geometry of Calabi--Yau hypersurfaces, let us briefly mention a few facts 
relating mainly to the structures of polyhedra.
The numbers of lattice points in the polytopes we found were of course bounded
by the corresponding minimal and maximal polytopes and ranged from 6 to 680.
The numbers of vertices ranged from 5 to 33.
An interesting object that we encountered is the 24-cell,
a self dual polytope with 24 vertices and 24 facets which 
arises as a subpolytope of the hypercube.
It is a Platonic solid that contains the Archimedian cuboctahedron 
(with symmetry order 48) as a reflexive section through the origin
parallel to one of its 24 bounding octahedra.
It has the maximal symmetry order $1152=24*48$ among all 4 dimensional
reflexive polytopes (in our context
symmetries are realized as lattice isomorphisms, i.e. as subgroups of
$GL(n,\IZ)$ and not as rotations).
The corresponding self mirror Calabi--Yau manifold has Hodge numbers (20,20).
The polytope with the largest order, namely 128, of 
$M_{\rm finest}/M_{\rm coarsest}$ is determined by the weight system 
$(1,1,1,1,4)/8$. For the 
Newton polytope of the quintic hypersurface in $\IP^4$, this order is 125.

For calculating the Hodge numbers of the Calabi--Yau hypersurfaces
corresponding to our reflexive polyhedra, we used the following well known 
formulae \cite{bat}:
\beq h_{11}=l(\D^*)-5-\sum_{{\rm codim }\th^* =1}l^*(\th^*)+
            \sum_{{\rm codim }\th^* =2}l^*(\th^*)l^*(\th)   \eeq
(this is also the Picard number) and 
\beq h_{12}=l(\D)-5-\sum_{{\rm codim }\th =1}l^*(\th)+
            \sum_{{\rm codim }\th =2}l^*(\th^*)l^*(\th),  \eeq
where $\th$ and $\th^*$ are faces of $\D$ and $\D^*$, respectively,
and $l(\cdot)$ and $l^*(\cdot)$ denote the numbers of integer points
and integer interior points of polytopes.
The result is a collection of 30,108 distinct pairs of Hodge numbers,
consisting of 14,986 `pairs of pairs' with $h_{11}\ne h_{12}$ and 136 pairs
with $h_{11}=h_{12}$, i.e. $\chi=2(h_{11}-h_{12})=0$.
These Hodge numbers are shown in the plot in appendix B. 
Because of the symmetry of the complete plot, we have only displayed the half
with $h_{11}\le h_{12}$.
While the general shape of this diagram has remained very similar to
corresponding plots of earlier lists of Hodge pairs \cite{CLS,KlS,nms,aas,k3},
there is a striking difference with regard to the density of points in the 
plot.
In particular, there is a large region, roughly bounded by $h_{11}\ge 15$,
$h_{12}\ge 15$ and $h_{11}+h_{12}\le 175$ in which every pair of numbers
actually occurs.

A peculiar feature of the complete plot is the existence of a few `isolated'
Hodge pairs in the lowest region. In particular, there are the pairs 
$\{21,1\}$, $\{20,5\}$, $\{19,7\}$ (these are the only ones with 
$h_{11}+h_{12}<28$) and $\{29,2\}$.
The Hodge pair $h_{11}=1$, $h_{12}=21$ is well known to correspond to a
quotient of the quintic hypersurface by a $\IZ_5$ symmetry
that acts without fixed points.
It is quite peculiar from the lattice point of 
view: Although the $N$ lattice is not the lattice $N_{\rm finest}$ generated 
by the vertices of $\D^*$, the only lattice points of $\D^*$ are its vertices
and the IP. 
Thus it provides an example where the $N$ lattice is not even generated
by the lattice points of $\D^*$. 
In terms of the geometric interpretation, this is necessary because modding
without fixed points means that no extra divisors and hence no extra lattice
points of $\D^*$ (except, possibly, points interior to facets) may be 
introduced.
In a similar way the model with Hodge numbers $(2,29)$ comes from modding the 
bicubic hypersurface in $\IP^2\times\IP^2$ by a fixed point free $\IZ_3$.
The hypersurface with Hodge numbers $(7,19)$ corresponds to a dual pair of
polytopes where $\D^*$ has 9 vertices and 12 lattice points and $\D$ has 10
vertices and 24 lattice points; for the model with Hodge numbers $(5,20)$ the 
corresponding numbers are 9, 10, 10 and 25, respectively.

\bigskip

{\it Acknowledgements:} 
We would like to thank the Computing Center at the TU Vienna, and in 
particular D.I. Ernst Haunschmid, for their helpful support.
M.K. is partly supported by the Austrian 
Research Funds FWF grant Nr. P11582-PHY. 
The research of H.S. is supported by the European Union 
TMR project ERBFMRX-CT-96-0045.

\newpage

\section*{Appendix A: r-minimal CWS and polytopes}

\ni\begin{tabular}{|r|rrrrr|c|} \hline 
d&$\!n_1$&$\!n_2$&$\!n_3$&$\!n_4$&$\!n_5$& \\ \hline\hline
5&1&1&1&1&1&1\\[-4pt]
6&1&1&1&1&2&2\\[-4pt]
7&1&1&1&2&2&-\\[-4pt]
7&1&1&1&1&3&-\\[-4pt]
8&1&1&2&2&2&2\\[-4pt]
8&1&1&1&2&3&-\\[-4pt]
8&1&1&1&1&4&3\\[-4pt]
9&1&1&2&2&3&-\\[-4pt]
9&1&1&1&3&3&1\\[-4pt]
9&1&1&1&2&4&-\\[-4pt]
10&1&1&2&3&3&-\\[-4pt]
10&1&1&2&2&4&-\\[-4pt]
10&1&1&1&3&4&-\\[-4pt]
10&1&1&1&2&5&-\\[-4pt]
11&1&1&2&3&4&-\\[-4pt]
11&1&1&2&2&5&-\\[-4pt]
11&1&1&1&3&5&-\\[-4pt]
12&2&2&2&3&3&-\\[-4pt]
12&1&1&3&3&4&1\\[-4pt]
12&1&1&2&4&4&-\\[-4pt]
12&1&1&2&3&5&-\\[-4pt]
12&1&1&1&4&5&-\\[-4pt]
12&1&1&2&2&6&1\\[-4pt]
12&1&1&1&3&6&-\\[-4pt]
13&1&1&3&4&4&-\\[-4pt]
13&1&1&3&3&5&-\\[-4pt]
13&1&1&2&4&5&-\\[-4pt]
13&1&1&2&3&6&-\\[-4pt]
13&1&1&1&4&6&-\\[-4pt]
14&2&2&3&3&4&-\\[-4pt]
14&1&1&3&4&5&-\\[-4pt]
14&1&1&2&4&6&-\\[-4pt]
14&1&1&2&3&7&-\\[-4pt]
14&1&1&1&4&7&-\\ \hline
\end{tabular}\hfill
\begin{tabular}{|r|rrrrr|c|} \hline 
d&$\!n_1$&$\!n_2$&$\!n_3$&$\!n_4$&$\!n_5$& \\ \hline\hline
15&2&2&3&3&5&-\\[-4pt]
15&1&1&3&5&5&-\\[-4pt]
15&1&1&3&4&6&-\\[-4pt]
15&1&1&2&5&6&-\\[-4pt]
15&1&1&3&3&7&-\\[-4pt]
15&1&1&2&4&7&-\\[-4pt]
15&1&1&1&5&7&-\\[-4pt]
16&1&1&4&5&5&-\\[-4pt]
16&1&1&4&4&6&-\\[-4pt]
16&1&1&3&5&6&-\\[-4pt]
16&1&1&3&4&7&-\\[-4pt]
16&1&1&2&5&7&-\\[-4pt]
16&1&1&3&3&8&-\\[-4pt]
16&1&1&2&4&8&1\\[-4pt]
16&1&1&1&5&8&-\\[-4pt]
17&1&1&4&5&6&-\\[-4pt]
17&2&2&3&3&7&-\\[-4pt]
17&1&1&3&5&7&-\\[-4pt]
17&1&1&3&4&8&-\\[-4pt]
17&1&1&2&5&8&-\\[-4pt]
18&1&1&4&6&6&-\\[-4pt]
18&1&1&4&5&7&-\\[-4pt]
18&1&1&3&6&7&-\\[-4pt]
18&2&2&3&3&8&-\\[-4pt]
18&1&1&3&5&8&-\\[-4pt]
18&1&1&2&6&8&-\\[-4pt]
18&1&1&3&4&9&-\\[-4pt]
18&1&1&2&5&9&-\\[-4pt]
18&1&1&1&6&9&-\\[-4pt]
19&1&1&4&6&7&-\\[-4pt]
19&1&1&3&6&8&-\\[-4pt]
19&1&1&3&5&9&-\\[-4pt]
19&1&1&2&6&9&-\\[-4pt]
20&2&2&5&5&6&-\\
\hline\end{tabular}\hfill
\begin{tabular}{|r|rrrrr|c|} \hline 
d&$\!n_1$&$\!n_2$&$\!n_3$&$\!n_4$&$\!n_5$& \\ \hline\hline
20&1&1&5&6&7&-\\[-4pt]
20&1&1&4&6&8&-\\[-4pt]
20&1&1&4&5&9&-\\[-4pt]
20&2&2&3&3&10&-\\[-4pt]
20&1&1&4&4&10&-\\[-4pt]
20&1&1&3&5&10&-\\[-4pt]
20&1&1&2&6&10&-\\[-4pt]
21&1&1&5&7&7&-\\[-4pt]
21&1&1&5&6&8&-\\[-4pt]
21&1&1&4&7&8&-\\[-4pt]
21&1&1&4&6&9&-\\[-4pt]
21&1&1&3&7&9&-\\[-4pt]
21&1&1&4&5&10&-\\[-4pt]
21&1&1&3&6&10&-\\[-4pt]
21&1&1&2&7&10&-\\[-4pt]
22&1&1&5&7&8&-\\[-4pt]
22&1&1&4&7&9&-\\[-4pt]
22&1&1&4&6&10&-\\[-4pt]
22&1&1&3&7&10&-\\[-4pt]
22&1&1&4&5&11&-\\[-4pt]
22&1&1&3&6&11&-\\[-4pt]
22&1&1&2&7&11&-\\[-4pt]
23&1&1&5&7&9&-\\[-4pt]
23&1&1&4&6&11&-\\[-4pt]
23&1&1&3&7&11&-\\[-4pt]
24&1&1&6&8&8&-\\[-4pt]
24&1&1&6&7&9&-\\[-4pt]
24&1&1&5&8&9&-\\[-4pt]
24&3&3&4&4&10&-\\[-4pt]
24&1&1&4&8&10&-\\[-4pt]
24&1&1&3&8&11&-\\[-4pt]
24&1&1&4&6&12&-\\[-4pt]
24&1&1&3&7&12&-\\[-4pt]
24&1&1&2&8&12&-\\
\hline\end{tabular}\\[-5mm]
\BC
Table 1a: r-minimal weight systems with $d\le 24$.
Here and in tables 2--7, the last column indicates the number of additional 
lattices on which $\D(\q)$ is r-maximal.

\ni\begin{tabular}{|r|rrrrr|} \hline 
d&$\!n_1$&$\!n_2$&$\!n_3$&$\!n_4$&$\!n_5$\\ \hline\hline
25&1&1&6&8&9\\[-4pt]
25&1&1&5&8&10\\[-4pt]
25&1&1&5&7&11\\[-4pt]
25&1&1&5&6&12\\[-4pt]
25&1&1&4&7&12\\[-4pt]
25&1&1&3&8&12\\[-4pt]
26&1&1&6&8&10\\[-4pt]
26&1&1&5&7&12\\[-4pt]
26&1&1&5&6&13\\[-4pt]
26&1&1&4&7&13\\[-4pt]
26&1&1&3&8&13\\[-4pt]
27&1&1&6&9&10\\[-4pt]
27&1&1&5&9&11\\[-4pt]
27&1&1&5&7&13\\[-4pt]
27&1&1&3&9&13\\[-4pt]
28&1&1&7&9&10\\[-4pt]
28&1&1&6&9&11\\[-4pt]
28&1&1&4&9&13\\[-4pt]
28&3&3&4&4&14\\[-4pt]
28&1&1&5&7&14\\[-4pt]
28&1&1&4&8&14\\[-4pt]
28&1&1&3&9&14\\[-4pt]
29&1&1&7&9&11\\[-4pt]
29&1&1&5&8&14\\[-4pt]
29&1&1&4&9&14\\[-4pt]
30&1&1&7&10&11\\[-4pt]
30&1&1&6&10&12\\[-4pt]
30&1&1&6&8&14\\[-4pt]
30&1&1&4&10&14\\[-4pt]
30&1&1&6&7&15\\[-4pt]
30&1&1&5&8&15\\[-4pt]
30&1&1&4&9&15\\[-4pt]
30&1&1&3&10&15\\[-4pt]
31&1&1&7&10&12\\[-4pt]
31&1&1&6&8&15\\
\hline\end{tabular}\hspace{5mm}
\begin{tabular}{|r|rrrrr|c|} \hline 
d&$\!n_1$&$\!n_2$&$\!n_3$&$\!n_4$&$\!n_5$\\ \hline\hline
31&1&1&4&10&15\\[-4pt]
32&1&1&8&10&12\\[-4pt]
32&1&1&6&8&16\\[-4pt]
32&1&1&5&9&16\\[-4pt]
32&1&1&4&10&16\\[-4pt]
33&1&1&8&11&12\\[-4pt]
33&1&1&7&11&13\\[-4pt]
33&1&1&6&9&16\\[-4pt]
33&1&1&4&11&16\\[-4pt]
34&1&1&8&11&13\\[-4pt]
34&1&1&6&9&17\\[-4pt]
34&1&1&4&11&17\\[-4pt]
35&1&1&7&9&17\\[-4pt]
35&1&1&5&11&17\\[-4pt]
36&1&1&9&12&13\\[-4pt]
36&1&1&8&12&14\\[-4pt]
36&1&1&5&12&17\\[-4pt]
36&1&1&7&9&18\\[-4pt]
36&1&1&6&10&18\\[-4pt]
36&1&1&5&11&18\\[-4pt]
36&1&1&4&12&18\\[-4pt]
37&1&1&9&12&14\\[-4pt]
37&1&1&7&10&18\\[-4pt]
37&1&1&5&12&18\\[-4pt]
38&1&1&7&10&19\\[-4pt]
38&1&1&5&12&19\\[-4pt]
39&1&1&9&13&15\\[-4pt]
39&1&1&5&13&19\\[-4pt]
40&1&1&10&13&15\\[-4pt]
40&1&1&8&10&20\\[-4pt]
40&1&1&7&11&20\\[-4pt]
40&1&1&5&13&20\\[-4pt]
41&1&1&8&11&20\\[-4pt]
42&1&1&10&14&16\\[-4pt]
42&1&1&6&14&20\\
\hline\end{tabular}\hspace{5mm}
\begin{tabular}{|r|rrrrr|c|} \hline 
d&$\!n_1$&$\!n_2$&$\!n_3$&$\!n_4$&$\!n_5$\\ \hline\hline
42&1&1&8&11&21\\[-4pt]
42&1&1&6&13&21\\[-4pt]
42&1&1&5&14&21\\[-4pt]
43&1&1&6&14&21\\[-4pt]
44&1&1&8&12&22\\[-4pt]
44&1&1&6&14&22\\[-4pt]
45&1&1&11&15&17\\[-4pt]
45&1&1&9&12&22\\[-4pt]
45&1&1&6&15&22\\[-4pt]
46&1&1&9&12&23\\[-4pt]
46&1&1&6&15&23\\[-4pt]
48&1&1&12&16&18\\[-4pt]
48&1&1&9&13&24\\[-4pt]
48&1&1&6&16&24\\[-4pt]
49&1&1&7&16&24\\[-4pt]
50&1&1&10&13&25\\[-4pt]
50&1&1&7&16&25\\[-4pt]
51&1&1&7&17&25\\[-4pt]
52&1&1&10&14&26\\[-4pt]
52&1&1&7&17&26\\[-4pt]
54&1&1&7&18&27\\[-4pt]
56&1&1&11&15&28\\[-4pt]
56&1&1&8&18&28\\[-4pt]
57&1&1&8&19&28\\[-4pt]
58&1&1&8&19&29\\[-4pt]
60&1&1&12&16&30\\[-4pt]
60&1&1&8&20&30\\[-4pt]
63&1&1&9&21&31\\[-4pt]
64&1&1&9&21&32\\[-4pt]
66&1&1&9&22&33\\[-4pt]
70&1&1&10&23&35\\[-4pt]
72&1&1&10&24&36\\[-4pt]
78&1&1&11&26&39\\[-4pt]
84&1&1&12&28&42\\[-4pt]
&&&&&\\
\hline\end{tabular}\\[2mm]
Table 1b: r-minimal weight systems with $d\ge 25$.
None of the corresponding polytopes is r-maximal on a different lattice.
\EC

\ni\begin{tabular}{|r|rrrrrr|c|} \hline 
d&$\!n_1$&$\!\!n_2$&$\!\!n_3$&$\!\!n_4$&$\!\!n_5$&$\!\!n_6$& \\ \hline\hline
4&1&1&1&1&0&0&\\[-7pt]4&1&1&0&0&1&1& 2 \\ \hline
4&1&1&1&1&0&0&\\[-7pt]5&1&2&0&0&1&1& - \\ \hline
4&1&1&1&1&0&0&\\[-7pt]6&1&1&0&0&2&2& - \\ \hline
4&1&1&1&1&0&0&\\[-7pt]6&2&2&0&0&1&1& - \\ \hline
4&1&1&1&1&0&0&\\[-7pt]6&1&3&0&0&1&1& - \\ \hline
4&1&1&1&1&0&0&\\[-7pt]7&2&3&0&0&1&1& - \\ \hline
4&1&1&1&1&0&0&\\[-7pt]8&2&4&0&0&1&1& - \\ \hline
4&1&1&1&1&0&0&\\[-7pt]9&3&4&0&0&1&1& - \\ \hline
4&1&1&1&1&0&0&\\[-7pt]10&3&5&0&0&1&1& - \\ \hline
4&1&1&1&1&0&0&\\[-7pt]12&4&6&0&0&1&1& - \\ \hline
5&1&2&1&1&0&0&\\[-7pt]5&1&2&0&0&1&1& - \\ \hline
5&1&2&1&1&0&0&\\[-7pt]5&2&1&0&0&1&1& - \\ \hline
5&1&2&1&1&0&0&\\[-7pt]6&2&2&0&0&1&1& - \\ \hline
5&1&2&1&1&0&0&\\[-7pt]6&1&3&0&0&1&1& - \\ \hline
5&1&2&1&1&0&0&\\[-7pt]6&3&1&0&0&1&1& - \\ \hline
5&1&2&1&1&0&0&\\[-7pt]7&2&3&0&0&1&1& - \\ \hline
5&1&2&1&1&0&0&\\[-7pt]7&3&2&0&0&1&1& - \\ \hline
\end{tabular}
\begin{tabular}{|r|rrrrrr|c|} \hline 
d&$\!n_1$&$\!\!n_2$&$\!\!n_3$&$\!\!n_4$&$\!\!n_5$&$\!\!n_6$& \\ \hline\hline
5&1&2&1&1&0&0&\\[-7pt]8&2&4&0&0&1&1& - \\ \hline
5&1&2&1&1&0&0&\\[-7pt]8&4&2&0&0&1&1& - \\ \hline
5&1&2&1&1&0&0&\\[-7pt]9&3&4&0&0&1&1& - \\ \hline
5&1&2&1&1&0&0&\\[-7pt]10&3&5&0&0&1&1& - \\ \hline
5&1&2&1&1&0&0&\\[-7pt]12&4&6&0&0&1&1& - \\ \hline
6&1&1&2&2&0&0&\\[-7pt]6&1&1&0&0&2&2& - \\ \hline
6&2&2&1&1&0&0&\\[-7pt]6&2&2&0&0&1&1& - \\ \hline
6&2&2&1&1&0&0&\\[-7pt]6&1&3&0&0&1&1& - \\ \hline
6&2&2&1&1&0&0&\\[-7pt]7&2&3&0&0&1&1& - \\ \hline
6&2&2&1&1&0&0&\\[-7pt]8&2&4&0&0&1&1& - \\ \hline
6&2&2&1&1&0&0&\\[-7pt]9&3&4&0&0&1&1& - \\ \hline
6&2&2&1&1&0&0&\\[-7pt]10&3&5&0&0&1&1& - \\ \hline
6&2&2&1&1&0&0&\\[-7pt]12&4&6&0&0&1&1& - \\ \hline
6&1&3&1&1&0&0&\\[-7pt]6&1&3&0&0&1&1& - \\ \hline
6&1&3&1&1&0&0&\\[-7pt]7&2&3&0&0&1&1& - \\ \hline
6&1&3&1&1&0&0&\\[-7pt]8&2&4&0&0&1&1& - \\ \hline
6&1&3&1&1&0&0&\\[-7pt]9&3&4&0&0&1&1& - \\ \hline 
\end{tabular}
\begin{tabular}{|r|rrrrrr|c|} \hline 
d&$\!n_1$&$\!\!n_2$&$\!\!n_3$&$\!\!n_4$&$\!\!n_5$&$\!\!n_6$& \\ \hline\hline
6&1&3&1&1&0&0&\\[-7pt]10&3&5&0&0&1&1& - \\ \hline
6&1&3&1&1&0&0&\\[-7pt]12&4&6&0&0&1&1& - \\ \hline
7&2&3&1&1&0&0&\\[-7pt]7&2&3&0&0&1&1& - \\ \hline
7&2&3&1&1&0&0&\\[-7pt]8&2&4&0&0&1&1& - \\ \hline
7&2&3&1&1&0&0&\\[-7pt]9&3&4&0&0&1&1& - \\ \hline
7&2&3&1&1&0&0&\\[-7pt]10&3&5&0&0&1&1& - \\ \hline
7&2&3&1&1&0&0&\\[-7pt]12&4&6&0&0&1&1& - \\ \hline
8&2&4&1&1&0&0&\\[-7pt]8&2&4&0&0&1&1& 1 \\ \hline
8&2&4&1&1&0&0&\\[-7pt]9&3&4&0&0&1&1& - \\ \hline
8&2&4&1&1&0&0&\\[-7pt]10&3&5&0&0&1&1& - \\ \hline
8&2&4&1&1&0&0&\\[-7pt]12&4&6&0&0&1&1& - \\ \hline
9&3&4&1&1&0&0&\\[-7pt]9&3&4&0&0&1&1& - \\ \hline
9&3&4&1&1&0&0&\\[-7pt]10&3&5&0&0&1&1& - \\ \hline
9&3&4&1&1&0&0&\\[-7pt]12&4&6&0&0&1&1& - \\ \hline
10&3&5&1&1&0&0&\\[-7pt]10&3&5&0&0&1&1& - \\ \hline
10&3&5&1&1&0&0&\\[-7pt]12&4&6&0&0&1&1& - \\ \hline
12&4&6&1&1&0&0&\\[-7pt]12&4&6&0&0&1&1& - \\ \hline 
\end{tabular}\\[-5mm]
\begin{center}
Table 2: r-minimal CWS of the type $\cont{\cont{\hbox{4+4}}}$.

\ni\begin{tabular}{|r|rrrrrr|c|} \hline 
d&$\!\!n_1$ &$\!\!n_2$&$\!\!n_3$&$\!\!n_4$&$\!\!n_5$&$\!\!n_6$&\\ \hline\hline
4&1&1&1&1&0&0&\\[-7pt]3&1&0&0&0&1&1& - \\ \hline
5&1&1&1&2&0&0&\\[-7pt]3&1&0&0&0&1&1& - \\ \hline
5&2&1&1&1&0&0&\\[-7pt]3&1&0&0&0&1&1& - \\ \hline
6&2&1&1&2&0&0&\\[-7pt]3&1&0&0&0&1&1& - \\ \hline
6&1&1&1&3&0&0&\\[-7pt]3&1&0&0&0&1&1& - \\ \hline
6&3&1&1&1&0&0&\\[-7pt]3&1&0&0&0&1&1& - \\ \hline
7&2&1&1&3&0&0&\\[-7pt]3&1&0&0&0&1&1& - \\ \hline
7&3&1&2&1&0&0&\\[-7pt]3&1&0&0&0&1&1& - \\ \hline
8&2&1&1&4&0&0&\\[-7pt]3&1&0&0&0&1&1& - \\ \hline
8&4&1&2&1&0&0&\\[-7pt]3&1&0&0&0&1&1& - \\ \hline
\end{tabular}
\begin{tabular}{|r|rrrrrr|c|} \hline 
d&$\!\!n_1$ &$\!\!n_2$&$\!\!n_3$&$\!\!n_4$&$\!\!n_5$&$\!\!n_6$&\\ \hline\hline
9&3&1&1&4&0&0&\\[-7pt]3&1&0&0&0&1&1& - \\ \hline
9&4&1&3&1&0&0&\\[-7pt]3&1&0&0&0&1&1& - \\ \hline
10&3&1&1&5&0&0&\\[-7pt]3&1&0&0&0&1&1& - \\ \hline
10&5&1&3&1&0&0&\\[-7pt]3&1&0&0&0&1&1& - \\ \hline
12&4&1&1&6&0&0&\\[-7pt]3&1&0&0&0&1&1& - \\ \hline
12&6&1&4&1&0&0&\\[-7pt]3&1&0&0&0&1&1& - \\ \hline
4&1&1&1&1&0&0&\\[-7pt]4&2&0&0&0&1&1& - \\ \hline
5&1&1&1&2&0&0&\\[-7pt]4&2&0&0&0&1&1& - \\ \hline
5&2&1&1&1&0&0&\\[-7pt]4&2&0&0&0&1&1& - \\ \hline
6&2&1&1&2&0&0&\\[-7pt]4&2&0&0&0&1&1& - \\ \hline
\end{tabular}
\begin{tabular}{|r|rrrrrr|c|} \hline 
d&$\!\!n_1$ &$\!\!n_2$&$\!\!n_3$&$\!\!n_4$&$\!\!n_5$&$\!\!n_6$&\\ \hline\hline
6&1&1&1&3&0&0&\\[-7pt]4&2&0&0&0&1&1& - \\ \hline
6&3&1&1&1&0&0&\\[-7pt]4&2&0&0&0&1&1& - \\ \hline
7&2&1&1&3&0&0&\\[-7pt]4&2&0&0&0&1&1& - \\ \hline
7&3&1&2&1&0&0&\\[-7pt]4&2&0&0&0&1&1& - \\ \hline
8&2&1&1&4&0&0&\\[-7pt]4&2&0&0&0&1&1& - \\ \hline
8&4&1&2&1&0&0&\\[-7pt]4&2&0&0&0&1&1& 1 \\ \hline
9&4&1&3&1&0&0&\\[-7pt]4&2&0&0&0&1&1& - \\ \hline
10&5&1&3&1&0&0&\\[-7pt]4&2&0&0&0&1&1& - \\ \hline
12&6&1&4&1&0&0&\\[-7pt]4&2&0&0&0&1&1& - \\ \hline
 & & & & & & &\\[-7pt] & & & & & & &  \\ \hline
\end{tabular}\\[6pt]
Table 3: r-minimal CWS of the type $\cont{\hbox{4+3}}$.

\ni
\begin{tabular}{|r|rrrrrr|c|} \hline 
d&$\!\!n_1$ &$\!\!n_2$&$\!\!n_3$&$\!\!n_4$&$\!\!n_5$&$\!\!n_6$&\\ \hline\hline
4 &1&1&1&1&0&0&\\[-7pt] 2&0&0&0&0&1&1& 2 \\ \hline
5 &1&1&1&2&0&0&\\[-7pt] 2&0&0&0&0&1&1& - \\ \hline
6 &1&1&2&2&0&0&\\[-7pt] 2&0&0&0&0&1&1& - \\ \hline
\end{tabular}
\begin{tabular}{|r|rrrrrr|c|} \hline 
d&$\!\!n_1$ &$\!\!n_2$&$\!\!n_3$&$\!\!n_4$&$\!\!n_5$&$\!\!n_6$&\\ \hline\hline
6 &1&1&1&3&0&0&\\[-7pt] 2&0&0&0&0&1&1& - \\ \hline
7 &1&1&2&3&0&0&\\[-7pt] 2&0&0&0&0&1&1& - \\ \hline
8 &1&1&2&4&0&0&\\[-7pt] 2&0&0&0&0&1&1& 1 \\ \hline
\end{tabular}
\begin{tabular}{|r|rrrrrr|c|} \hline 
d&$\!\!n_1$ &$\!\!n_2$&$\!\!n_3$&$\!\!n_4$&$\!\!n_5$&$\!\!n_6$&\\ \hline\hline
9 &1&1&3&4&0&0&\\[-7pt] 2&0&0&0&0&1&1& - \\ \hline
10&1&1&3&5&0&0&\\[-7pt] 2&0&0&0&0&1&1& - \\ \hline
12&1&1&4&6&0&0&\\[-7pt] 2&0&0&0&0&1&1& - \\ \hline
\end{tabular}\\[6pt]
Table 4: r-minimal CWS of the type 4+2. 

\ni
\begin{tabular}{|r|rrrrrr|c|} \hline 
d&$\!\!n_1$ &$\!\!n_2$&$\!\!n_3$&$\!\!n_4$&$\!\!n_5$&$\!\!n_6$&\\ \hline\hline
3 &1&1&1&0&0&0&\\[-7pt] 3&0&0&0&1&1&1& 1 \\ \hline
\end{tabular}
\begin{tabular}{|r|rrrrrr|c|} \hline 
d&$\!\!n_1$ &$\!\!n_2$&$\!\!n_3$&$\!\!n_4$&$\!\!n_5$&$\!\!n_6$&\\ \hline\hline
3 &1&1&1&0&0&0&\\[-7pt] 4&0&0&0&2&1&1& - \\ \hline
\end{tabular}
\begin{tabular}{|r|rrrrrr|c|} \hline 
d&$\!\!n_1$ &$\!\!n_2$&$\!\!n_3$&$\!\!n_4$&$\!\!n_5$&$\!\!n_6$&\\ \hline\hline
4 &2&1&1&0&0&0&\\[-7pt] 4&0&0&0&2&1&1& 1 \\ \hline
\end{tabular}\\[6pt]
Table 5: r-minimal CWS of the type 3+3. 
\EC

\ni
\begin{minipage}{9cm}
\begin{center}
\begin{tabular}{|r|rrrrrrr|c|} \hline 
d&$n_1$ &$n_2$ &$n_3$ &$n_4$ &$n_5$ & $n_6$ &$n_7$ & \\ \hline\hline
3&1&1&1&0&0&0&0&\\[-7pt]3&1&0&0&1&1&0&0&-\\[-7pt]3&1&0&0&0&0&1&1&\\ \hline
3&1&1&1&0&0&0&0&\\[-7pt]3&1&0&0&1&1&0&0&-\\[-7pt]4&2&0&0&0&0&1&1&\\ \hline
3&1&1&1&0&0&0&0&\\[-7pt]4&2&0&0&1&1&0&0&-\\[-7pt]4&2&0&0&0&0&1&1&\\ \hline
4&2&1&1&0&0&0&0&\\[-7pt]4&2&0&0&1&1&0&0&1\\[-7pt]4&2&0&0&0&0&1&1&\\ \hline 
\hline
3&1&1&1&0&0&0&0&\\[-7pt]3&1&0&0&1&1&0&0&-\\[-7pt]2&0&0&0&0&0&1&1&\\ \hline
3&1&1&1&0&0&0&0&\\[-7pt]4&2&0&0&1&1&0&0&-\\[-7pt]2&0&0&0&0&0&1&1&\\ \hline
4&2&1&1&0&0&0&0&\\[-7pt]4&2&0&0&1&1&0&0&1\\[-7pt]2&0&0&0&0&0&1&1&\\ \hline
\hline
3&1&1&1&0&0&0&0&\\[-7pt]2&0&0&0&1&1&0&0&-\\[-7pt]2&0&0&0&0&0&1&1&\\ \hline
4&2&1&1&0&0&0&0&\\[-7pt]2&0&0&0&1&1&0&0&1\\[-7pt]2&0&0&0&0&0&1&1&\\ \hline
\end{tabular}\\[6pt]
Table 6: r-minimal CWS of the types\\ 
$\cont{\hbox{3+3}\hskip -3pt}\hskip -2pt\cont{\hbox{~+3}}$, 
$\cont{\hbox{3+3}}$+2 and 3+2+2.
\\[7mm]
\begin{tabular}{|r|rrrrrrrr|c|} \hline 
d &$\!n_1$&$\!n_2$&$\!n_3$&$\!n_4$&$\!n_5$&$\!n_6$&$\!n_7$&$\!n_8$&\\ 
\hline\hline
2&1&1&0&0&0&0&0&0& \\[-7pt]2&0&0&1&1&0&0&0&0& \\[-7pt]
2&0&0&0&0&1&1&0&0&1\\[-7pt]2&0&0&0&0&0&0&1&1& \\ \hline 
\end{tabular}
\\[6pt]
Table 7: The r-minimal CWS of the\\ 
type 2+2+2+2. 
\end{center}
\end{minipage}
\hfill
\begin{minipage}{7cm}
\begin{center}
\begin{tabular}{|rrrrrrr|} \hline
$v_1$&$v_2$&$v_3$&$v_4$&$v_5$&$v_6$&$v_7$\\ \hline\hline
1&0&-2&0&0&2&\\[-7pt]
0&1&-1&0&0&0&\\[-7pt]
0&0&0&1&0&-1&\\[-7pt]
0&0&0&0&1&-1&\\ \hline
0&1&1&1&1&1&\\ \hline\hline
1&0&0&-2&0&-3&\\[-7pt]
0&1&1&-3&0&-4&\\[-7pt]
0&0&2&-3&0&-4&\\[-7pt]
0&0&0&0&1&-1&\\ \hline
3&2&2&0&1&1&\\ \hline\hline
1&-1&0&0&-2&0&2\\[-7pt]
0&0&1&0&-1&0&-1\\[-7pt]
0&0&0&1&-1&0&0\\[-7pt]
0&0&0&0&0&1&-1\\ \hline
0&0&2&1&1&1&1\\ \hline\hline
1&0&-1&0&-2&0&-1\\[-7pt]
0&1&1&0&-2&0&2\\[-7pt]
0&0&0&1&-1&0&0\\[-7pt]
0&0&0&0&0&1&-1\\ \hline
3&0&0&1&1&1&1\\ \hline\hline
1&0&-1&0&-1&0&-2\\[-7pt]
0&1&1&0&2&0&-4\\[-7pt]
0&0&0&1&-1&0&0\\[-7pt]
0&0&0&0&0&1&-1\\ \hline
4&0&0&2&2&1&1\\ \hline\hline
1&1&-3&0&-4&0&1\\[-7pt]
0&2&-2&0&-3&0&3\\[-7pt]
0&0&0&1&-1&0&0\\[-7pt]
0&0&0&0&0&1&-1\\ \hline
3&0&0&1&1&1&1\\ \hline\hline
1&1&-3&0&-5&0&1\\[-7pt]
0&2&-2&0&-3&0&3\\[-7pt]
0&0&0&1&-1&0&0\\[-7pt]
0&0&0&0&0&1&-1\\ \hline
4&0&0&1&1&1&1\\ \hline
\end{tabular}\\[6pt]
Table 8: The 7 last r-minimal\\
polytopes: Vertices, weights
\end{center}
\end{minipage}

\newpage

\section*{Appendix B: The Hodge--plot}

\vspace{1cm}

\epsfbox{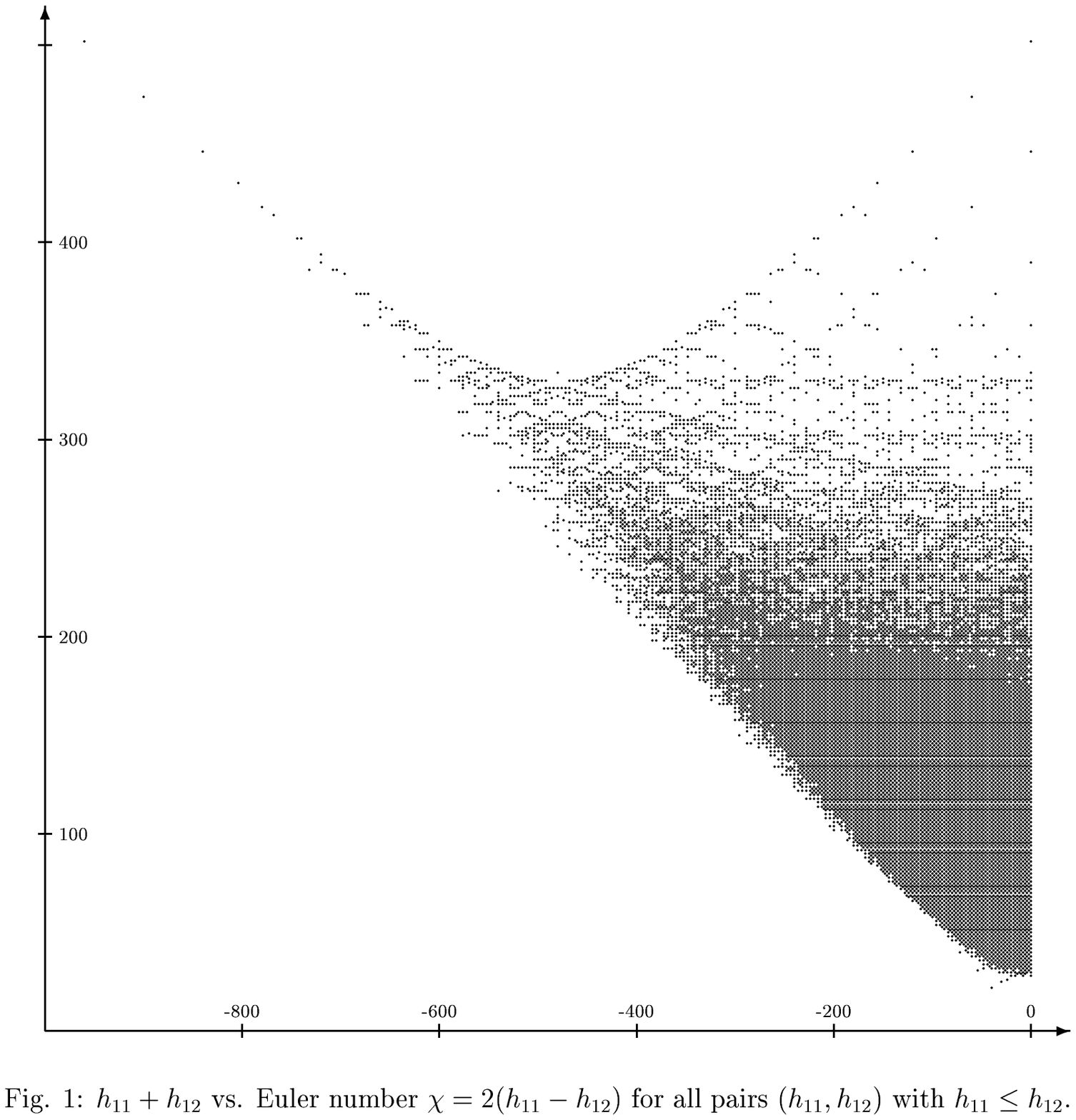}

\end{document}